\documentclass[aps,prc,twocolumn]{revtex4}
\usepackage{graphicx}
\usepackage{amsmath,bbm}
\usepackage{amssymb,bm}
\usepackage{array}

\begin{document}
\title{Upsilon suppression at energies available at the BNL Relativistic Heavy Ion Collider and at the CERN Large Hadron Collider in a modified color screening scenario}
\author{P.~K.~Srivastava\footnote{$prasu111@gmail.com$}$^1$}
\author{S. K. Tiwari$^1$}
\author{C.~P.~Singh$^1$}

\affiliation{$^1$Department of Physics, Banaras Hindu University, 
Varanasi 221005, INDIA}

\begin{abstract}

The suppression of heavy quarkonia e.g. $J/\psi$, $\Upsilon$ etc. is considered as a suitable probe to identify the nature of the matter created in heavy ion collisions. Recently we have presented a modified colour screening model for $J/\psi$ suppression in the quark gluon plasma (QGP) using quasiparticle model as the equation of state. In this paper, we extend our model to calculate the anomalous suppression of various states of $\Upsilon$ arising due to QGP medium alone. We obtain the suppression patterns of the different bottomonia states with respect to centrality at various available collision energies and compare them with the available experimental data.
\\

 PACS numbers: 12.38.Mh, 12.38.Gc, 25.75.Nq, 24.10.Pa
\end{abstract}

\maketitle 
\section{Introduction}
\noindent
The prediction of a possible existence of a deconfined quark gluon plasma (QGP) phase at high temperature and/or density by quantum chromodynamics (QCD) opened up a new challanging task of detection of this deconfined state of strongly interacting matter~\cite{singh}. It seems worth exploring the nature and properties of this novel state by proposing some suitable probes. Heavy quarkonia are expected to play an important role in testing QCD and investigating the nature of the QCD phase of quarks and gluons~\cite{bram}. Heavy quarkonia ($J/\psi$, $\Upsilon$ etc.) suppressions have long been considered as a clean signal for deconfining phase transition and QGP formation in the heavy ion collision experiments~\cite{mats}. The idea relies mainly on colour screening mechanism of heavy quark potential similar to the electric charge screening in QED plasma~\cite{shuryak}. However, the anomalaous suppression cannot be assigned to Debye screening alone in QGP~\cite{strick}. Complications arise from the other factors like the existence of Landau damping of heavy quark potential~\cite{ima1,ima2}, the non-perturbative effects involved in the study of QCD phase transition from hadron gas (HG) to QGP and the suppression of heavy quarkonia from conventional nuclear effects collectively known as cold nuclear matter (CNM) effects~\cite{vogt} eg., nuclear absorption, shadowing and anti-shadowing in parton distribution functions, etc. Further it is observed that charmonia states e.g. $J/\psi,~\chi_{c}$ etc. are not very suitable probes because significant contributions arise from the regeneration effects also at higher energies especially at LHC where statistical production of $c$ and $\bar{c}$ becomes large~\cite{grand,grand1}. In comparison to charmonia, the suppression pattern of upsilon ($\Upsilon$) and its excited states is regarded as cleaner signals to study the properties of the medium created at high temperature and/or density. The factors responsible for this fact are large masses and small binding radii of various bottomonia states. Due to large masses, the probability of regeneration of bottomonia by possible coalescence of $b-\bar{b}$ pair is expected to be almost negligible in comparison to charmonia states. However, a recent work by Emerick and collaborators have shown that the regeneration can still play an important role in the upsilon production~\cite{emerick}. The magnitude of the contribution depends on an interplay of the masses of the open- and hidden-bottom states in a system with fixed $b\bar{b}$ content. Further, the smallness of the ratio of hidden- to open-bottom states in elementary collisions implies that even small contributions to bottomonium regeneration can be significant relative to primordial production. In Ref.~\cite{emerick}, the authors have considered the temperature dependence of bottomonia binding energies in weak-binding scenario (WBS) and/or strong-binding scenario (SBS). In weak binding scenario, the effect of regeneration through recombination is very less. However, it becomes significant in the case of SBS~\cite{emerick}. 

The major hurdle in estimating precisely the anomalous suppression of quarkonia involves the almost unknown background contribution as discussed previously. It is argued recently that if we take the ratio of the yields of different bottomonia states, then CNM effects should almost cancel out in the ratio~\cite{vogt1} and one can get the amount of suppression arising solely due to QGP. Thus, the suppression pattern of various states of $\Upsilon$ can provide an important insight into the properties of the medium created in heavy ion collisions.

Bottomonia suppressions in heavy-ion collisions have experimentally been studied at Relativistic Heavy Ion Collider (RHIC) by STAR~\cite{rosi} and PHENIX~\cite{Adare} experiments at $\sqrt{s_{NN}}=200$ GeV. However, due to limited vertex resolution, they are not able to disentangle different bottomonia states ie., $\Upsilon (1S)$, $\Upsilon (2S)$, and $\Upsilon (3S)$. Thus STAR only provided the nuclear modification factor for the combined production of different states ie., $\Upsilon (1S)+\Upsilon (2S)+\Upsilon (3S)$, with respect to the number of participants ($N_{part}$)~\cite{rosi}. Recently CMS detector at Large Hadron Collider (LHC) has measured the suppression patterns of different upsilon states separately. The CMS collaboration reported their initial measurements for the absolute $\Upsilon(1S)$ suppression as well as the relative suppression for $\Upsilon (2S)+\Upsilon (3S)$ with respect to $\Upsilon (1S)$ and find that the excited states $\Upsilon(nS)$ are suppressed with respect to $\Upsilon (1S)$~\cite{cms1,cms2}. Further they have also presented the suppression patterns of different states of upsilon separately with a larger data set and show their relative ratios also and finally observed the sequential suppression of bottomonia states~\cite{cms3}. They have also plotted the ratios of nuclear modification factors of different upsilon states~\cite{cms3} to illustrate the anomalous suppression arising due to QGP without any CNM effect as suggested by earlier studies~\cite{vogt1}. However, a vigorous experimental effort to quantify CNM effects is still continuing at RHIC~\cite{cnm}.

We have recently modified the colour-screening model of Chu and Matsui~\cite{chu} by following two steps. Firstly we parametrize the pressure~\cite{mmish} instead of energy density as used by them since pressure density becomes almost zero at the deconfining phase transition point. Secondly, we employ the quasiparticle description instead of bag model equation of state (EOS) for QGP~\cite{prashant}. This highlights a major difference between our present approach and the models of Chu, Matsui~\cite{chu} or Mishra et al~\cite{mmish} because bag model often gives a crude EOS for QGP. Feed down from higher resonances have also been incorporated in the model. We have earlier used this model to calculate the survival probability of $J/\psi$ with respect to number of participant ($N_{part}$) involved in the collision. We surprisingly find that our model reproduces the CNM normalized data for $J/\psi$ suppression at all the energies including CERN SPS, BNL RHIC, and CERN LHC energies~\cite{prashant}. The suppression pattern of $J/\psi$ in the central collision has a complicated pattern of going from $0.75$ at SPS, to $0.4$ at RHIC and is back up to $0.6$ at LHC. Less suppression for $J/psi$ at LHC in comparison to RHIC has been observed mainly due to the $p_{T}$-range used in the integration. As such experimental $p_{T}$-range varies between $0.3$ to $7$ GeV for RHIC data while it varies from $6.5$ GeV to $30$ GeV in the case of LHC data. However, fo SPS this range lies between $0.1$ and $5$ GeV. We have taken the same $p_{T}$-range in our model as was used in experimental data. Charmonia involving large momentum will be formed at a later stage as seen in the plasma rest frame. Consequently the region covered by a hot plasma is thus reduced and we thus expect less suppression for such states~\cite{karsch}. Larger momentum at LHC makes the survival probability of quarkonia states larger in comparison to  the survival probability at RHIC for most central collisions. In addition to the above factor, we should also consider the variation in the energy density which is larger at LHC and hence this causes more dissociations. 

  In this paper, our motivation is to extend our model for explaining the suppressions of various bottomonia states. Since the suppression of bottomonia due to cold nuclear matter (CNM) is not precisely determined, our calculation can be useful to provide an indirect estimate of CNM effect. We will also calculate the ratio of suppression pattern of various states to see whether the effects of cold nuclear matter in the ratios have any cancellation effect. Further, we also calculate the survival probability for $\Upsilon (1S)$, and $\Upsilon (2S)$ with respect to transverse momentum $p_{T}$ in order to illustrate the success of our model in explaining the experimental data. 
\section{Formulation}
\subsection{Cooling law}
We have used the colour screening idea of Chu and Matsui~\cite{chu}. However, instead of using bag model for QGP, we now use quasiparticle model (QPM) as new EOS of QGP. We assume that the QGP medium formed during the collision, expands and cools according to the Bjorken's boost invariant longitudinal viscous hydrodynamics in mid-rapidity region. Employing the conservation of energy-momentum tensor, the rate of the decrease of energy density $\epsilon $~\cite{teany} is given by
\begin{equation}
\frac{d\epsilon}{d\tau}=-\frac{\epsilon+p}{\tau}+\frac{4\eta}{3\tau^2},
\end{equation}
where $\eta$ is the shear viscosity of the QGP medium, $p$ is the pressure and $\tau$ represents the proper time. The energy density and pressure are computed by using QPM EOS~\cite{pks,pks1} for QGP. Using Eq.(1) and the thermodynamical identity $\epsilon=T\frac{dp}{dT}-p$, the cooling laws for energy density and pressure in the QPM model can be separately given as~\cite{prashant}:
\begin{equation}
\epsilon=c_1+c_2\tau^{-q}+\frac{4\eta}{3c_s^2}\frac{1}{\tau},
\end{equation}

\begin{equation}
p=-c_1+c_2\frac{c_s^2}{\tau^q}+\frac{4\eta}{3\tau}\left(\frac{q}{c_s^2-1}\right)+c_3\tau^{-c_s^2},
\end{equation}
where $c_1$ , $c_2$ and $c_3$ are constants which can be determined by imposing the initial boundary conditions on energy density and pressure, $q=c_{s}^{2}+1$ with $c_{s}$ being the speed of sound in the medium. We take $\epsilon=\epsilon_0$ at $\tau=\tau_0$ (initial thermalization time) and also $\epsilon=0$ at $\tau=\tau^{'}$; where $\tau^{'}$ is the proper time. Consequently, the constants $c_1$ and $c_2$ are given as~\cite{prashant} :
\begin{equation}
c_1=-c_2\tau '^{-q}-\frac{4\eta}{3c_s^2\tau^{'}}
\end{equation}
, where $\tau^{'}=\tau_0 A^{-\frac{3R}{R-1}}$, $A=T_0/T^{'}$ and $R$ is the Reynold's number for QGP. Further :
\begin{equation}
c_2=\frac{\epsilon_0-\frac{4\eta}{3c_s^2}\left(\frac{1}{\tau_0}-\frac{1}{\tau^{'}}\right)}{\tau_0^{-q}-\tau^{'-q}}.
\end{equation}
Using the initial condition for $p=p_0$ at $\tau=\tau_0$, we find the value of $c_3$ as~\cite{prashant} :

\begin{equation}
c_3=(p_0+c_1)\tau_0^{c_s^2}-c_2c_s^2\tau_0^{-1}-\frac{4\eta}{3}\left(\frac{q}{c_s^2-1}\right)\tau_0^{(c_s^2-1)}.
\end{equation}
\subsection{Pressure Profile}
We take a pressure profile function in the transverse plane with a transverse distance $r$ as~\cite{mmish,prashant} : 
\begin{equation}
p(t_i,r)=p(t_i,0)h(r); \quad h(r)=\left(1-\frac{r^2}{R_T^2}\right)^{\beta}\theta(R_T-r),
\end{equation} 
where the coefficient $p(t_i,0)$ is yet to be determined, $R_T$ denotes the radius of the cylinderical plasma and it is related to the transverse overlap area $A_T$ as determined by Glauber model $R_T=\sqrt{\frac{A_T}{\pi}}$~\cite{balver,adler}. The pressure is thus assumed to be maximum at the central axis but it vanishes at the edge $R_{T}$ where hadronization first begins. The exponent $\beta$ depends on the energy deposition mechanism and here we have taken $\beta=1.0$~\cite{mmish,prashant}; $\theta$ is the unit step-function. The factor $p(t_i,0)$ is related to the average initial pressure $<p>_i$~\cite{prashant} :
\begin{equation}
p(t_i,0)=(1+\beta)<p>_i.
\end{equation} 
The average pressure is determined by the centrality dependent initial average energy density $<\epsilon>_i$ which is further given by Bjorken's formula~\cite{adler,bjor} :
\begin{equation}
<\epsilon>_i=\frac{1}{A_T\tau_i} \frac{dE_T}{dy}. 
\end{equation}
Here $dE_T/dy$ is the transverse energy deposited per unit rapidity. We use the experimental value of $dE_{T}/d\eta^{'}$ where $\eta^{'}$ is pseudorapidity and then multiply it by a corresponding Jacobian factor~\cite{adler,cms} to obtain $dE_{T}/dy$ for a given number of participants ($N_{part}$) at a particular center-of-mass energy ($\sqrt{s_{NN}}$). At the initial proper time, $\frac{\partial <p>_{i}}{\partial <\epsilon>_{i}}=\frac{<p>_{i}}{<\epsilon>_{i}}=c_{s}^{2}$ as given by EOS of QGP in QPM~\cite{pks,pks1} and thus $<p>_{i}=c_{s}^{2}<\epsilon>_{i}$.

\subsection{Constant Pressure Contour and Radius of Screening Region} 
Since the cooling law for pressure cannot be solved for $\tau$ and, therefore, we use a trick to determine the radius of screening region. Writing the cooling law of pressure as follows~\cite{prashant} :

\begin{equation}
p(\tau,r)=A+\frac{B}{\tau^{q}}+\frac{C}{\tau} +\frac{D}{\tau^{c_s^2}},
\end{equation}
where $A$, $B$, $C$ and $D$ are constants related to $c_1$, $c_2$ and $c_3$ as : $A=-c_1$, $B=c_2 c_s^2$, $C=\frac{4\eta q}{3(c_s^2-1)}$ and $D=c_3$. Writing the above equation at $\tau=\tau_i$ and at screening time $\tau=\tau_s$ we get : 
\begin{equation}
p(\tau_i,r)=A+\frac{B}{\tau_i^{q}}+\frac{C}{\tau_i}+\frac{D}{\tau_i^{c_s^2}}=p(\tau_{i},0)h(r),
\end{equation} 
and
\begin{equation}
p(\tau_s,r)=A+\frac{B}{\tau_s^{q}}+\frac{C}{\tau_s}+\frac{D}{\tau_s^{c_s^2}}=~p_{QGP}.
\end{equation}
Here $p_{QGP}$ is the QGP pressure as determined by EOS in QPM~\cite{pks}. Solving Eqs. (11) and (12) numerically and equating the screening time $\tau_s$ to the dilated formation time of quarkonia $t_{F}$ (=$\gamma \tau_F$ where $\gamma=E_{T}/M_{\Upsilon}$ is the Lorentz factor associated with the transverse motion of the $b-\bar{b}$ pair, $M_{\Upsilon}=9.46$ GeV and $\tau_{F}$ is the proper time required for $b-\bar{b}$ pair in the formation of $\Upsilon$~\cite{pkshukla,karsch1}), we can find the radius of the screening region $r_{s}$. The screening region involves the temperature larger than the dissociation temperature so that the quarkonia formation becomes unlikely inside that region~\cite{mmish}. Hence the pair will in all probability escape from the screening region and form the quarkonia if $|\vec r_{\Upsilon}+\vec v t_F|\ge r_s$ where $\vec r_{\Upsilon}$ is the position vector at which the charm-quark pair is created~\cite{chu,mmish}.

The above kinematic condition takes a simplified form by assuming that $\Upsilon$ is moving with transverse momentum $p_T$. Thus the above escape condition can be expressed in the form of a trigonometric condition~\cite{mmish,prashant} :

\begin{equation}
\cos \phi\ge Y ;\quad  Y=\frac{(r_s^2-r_{\Upsilon}^2)m-\tau_F^2p_T^2/m}{2r_{\Upsilon}\tau_F p_T},
\end{equation}
where $\phi$ is the angle between the transverse momentum ($p_{T}$) and the position vector $\vec r_{\Upsilon}$ and $r_{\Upsilon}=|\vec r_{\Upsilon}|$ with $m=M_{\Upsilon}$.
\subsection{Survival Probability}
Assuming the radial probability distribution for the production of $b\bar b$ pair in hard collisions at transverse distance $r$ as 
\begin{equation}
f(r)\propto\left(1-\frac{r^2}{R_T^2}\right)^{\alpha}\theta(R_T-r).
\end{equation}    
Here we take $\alpha=1/2$ in our calculation as used in Ref.~\cite{chu}. Then, in the colour screening scenario, the survival probability for the quarkonia can easily be obtained as~\cite{mmish,chu} :

\begin{equation}
S(p_T,N_{part})=\frac{2(\alpha+1)}{\pi R_T^2}\int_0^{R_T}dr r \phi_{max}(r)\left\{1-\frac{r^2}{R_T^2}\right\}^{\alpha},
\end{equation}
where the maximum positive angle $\phi_{max}$ allowed by Eq. (13) becomes~\cite{prashant} :
$$
\phi_{max}(r)=\left\{\begin{array}{rl}
\pi     & \mbox{~~if $Y\le -1$}\\
\pi-\cos^{-1}|Y|  & \mbox{~~if $0\ge Y\ge -1$}\\
\cos^{-1}|Y| & \mbox{~~$0\le Y\le -1$}\\
 0        & \mbox{~~$Y \ge 1$} 
\end{array}\right.
$$
since the experimentalists always measure the quantity namely $p_T$ integrated nuclear modification factor. We get the theoretical $p_T$ integrated survival probability as follows :

\begin{equation}
S (N_{part})=\frac{\int_{p_{Tmin}}^{p_{Tmax}}S(p_T,N_{part})dp_T}{\int_{p_{Tmin}}^{p_{Tmax}}d p_T}.
\end{equation} 
    
It has been found that only about $50.9\%$ of the observed $\Upsilon(1S)$ come from hard collisions whereas $10.7\%$, $0.8\%$, $27.1\%$ and $10.5\%$is from the decay of $\Upsilon(2S)$, $\Upsilon(3S)$, $\chi_{b1}$ and $\chi_{b2}$, respectively~\cite{b}. Therefore, the net survival probability of $\Upsilon(1S)$ in the presence of QGP medium is :

\begin{eqnarray}
S_{\Upsilon(1S)} &=& 0.509\langle S_{\Upsilon(1S)}\rangle_{p_T}+0.107 \langle S_{\Upsilon(2S)}\rangle_{p_T}\\ \nonumber
&+& 0.008 \langle S_{\Upsilon(3S)}\rangle_{p_T}+0.271 \langle S_{\chi_{b1}}\rangle_{p_T}\\ \nonumber
&+& 0.105\langle S_{\chi_{b2}}\rangle_{p_T}.
\end{eqnarray}
Similarly, the net survival probability of $\Upsilon(2S)$ can be obtained in QGP as~\cite{mic}: 
\begin{equation}
S_{\Upsilon(2S)}= 0.5\langle S_{\Upsilon(2S)}\rangle_{p_T}+0.5\langle S_{\Upsilon(3S)}\rangle_{p_T}
\end{equation} 
\begin{table}
\caption{Values of the parameters.}
\begin{tabular}{l|l|l|l|l|l}
\hline\hline
~  & $T_i(GeV)$ & $p_i(GeV^4)$ & $s_i(GeV^3)$ & $\alpha$ & $\beta$\\\hline
SPS & 0.5 & 0.25 & 2.009 & 0.5 & 1.0 \\\hline
RHIC & 0.5 & 0.25 & 2.009 & 0.5 & 1.0\\\hline
LHC & 1.0 & 4.5 & 16.41 & 0.5 & 1.0\\
\hline\hline
\end{tabular}
\end{table}

\begin{table}
\caption{Masses, formation times and dissociation temperatures of the quarkonia.}
\begin{tabular}{l|l|l|l|l}
\hline\hline
    ~    & $m$(GeV) & $\tau_F$(fm) & Set I  ($T_D/T_c$) & Set II ($T_D/T_c$)\\\hline
$\Upsilon(1S)$ & 9.46 & 0.76 & 4.0 & 2.0 \\\hline
$\Upsilon(2S)$ & 10.02 & 1.9 & 1.6  & 1.2 \\ \hline
$\Upsilon(3S)$ & 10.36 & 2.0 & 1.17 & 1.0 \\ \hline
$\chi_{b1}$    & 9.99 & 2.6 & 1.76 & 1.3 \\ \hline
$\chi_{b2}$    & 10.26 & 2.6 & 1.19 & 1.19 \\
\hline\hline
\end{tabular}
\end{table}
In our calculation, we use $T_{c}=0.17$ GeV in accordance with the recent lattice QCD results~\cite{borsnyi}. We use the initial thermalization time ($\tau_{i}$) as $0.5$ fm. Similarly initial tempertaure ($T_{i}$), pressure density ($p_{i}$), entropy density ($s_{i}$) at proper time $\tau_{i}$ along with $\alpha$ and $\beta$ at different energies are tabulated in Table. 1. The value of $T_{i}$, $p_{i}$ and $s_{i}$ are taken in accordance with our QPM results~\cite{pks}. Other parameters like masses ($m$), formation time ($\tau_{F}$) and two different sets of dissociation temperatures ($T_{D}/T_{c}$) labelled as set I and set II for different quarkonia states are given in Table. 2~\cite{satz,mocsy}. The reason behind the use of two different sets of dissociation temperature will be discussed later. 


\begin{figure}[!ht]
    \centering
        \includegraphics[height=18em]{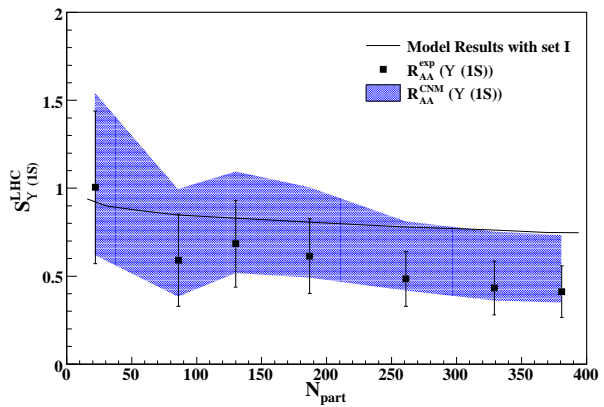}
    \label{fig:Fig 2}
    \caption{(Colour online) Variation of survival probability of $\Upsilon(1S)$ with respect to number of participants ($N_{part}$) using dissociation temperature of set I at LHC energy ($\sqrt{s_{NN}}=2.76$ TeV). Solid squares are the experimental data points without CNM normalization~\cite{cms3}. Blue band presents the $R_{AA}^{CNM}$ which is obtained from $R_{AA}^{exp}~/~S_{\Upsilon(1S)}^{LHC}$ }
\end{figure}

\begin{figure}[!ht]
    \centering
        \includegraphics[height=18em]{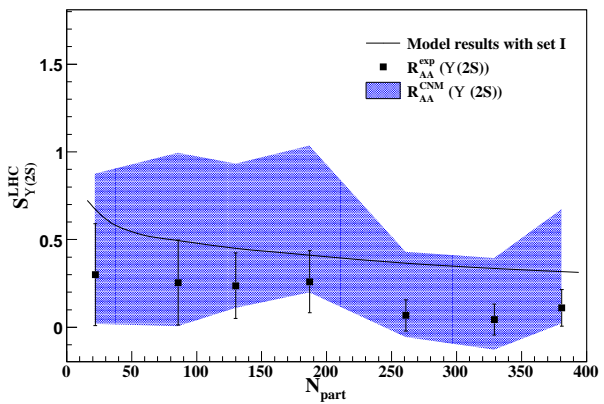}
    \label{fig:Fig 4}
    \caption{(Colour online) Variation of survival probability of $\Upsilon(2S)$ with respect to number of participants ($N_{part}$) using dissociation temperature of set I at LHC energy ($\sqrt{s_{NN}}=2.76$ TeV). Solid squares are the experimental data points without CNM normalization~\cite{cms3}. Blue band presents $R_{AA}^{CNM}$ for $\Upsilon(2S)$.}
\end{figure}
\begin{figure}[!ht]
    \centering
        \includegraphics[height=18em]{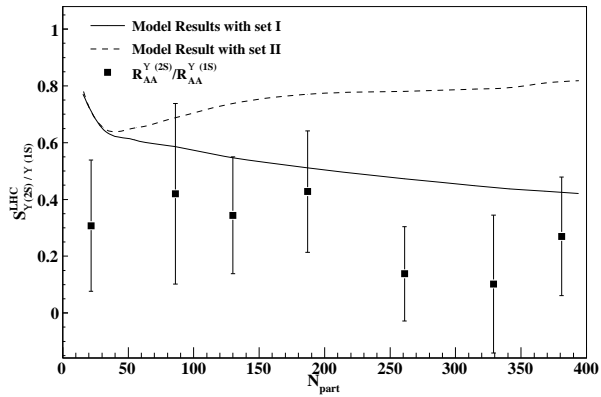}
    \label{fig:Fig 5}
    \caption{Variation of survival probability of $\Upsilon(2S)/\Upsilon(1S)$ with respect to number of participants ($N_{part}$) using dissociation temperature of set I (solid curve) and set II (dashed-curve) at LHC energy ($\sqrt{s_{NN}}=2.76$ TeV). Solid squares are the experimental data points~\cite{cms3}.}
\end{figure}

\begin{figure}[!ht]
    \centering
        \includegraphics[height=18em]{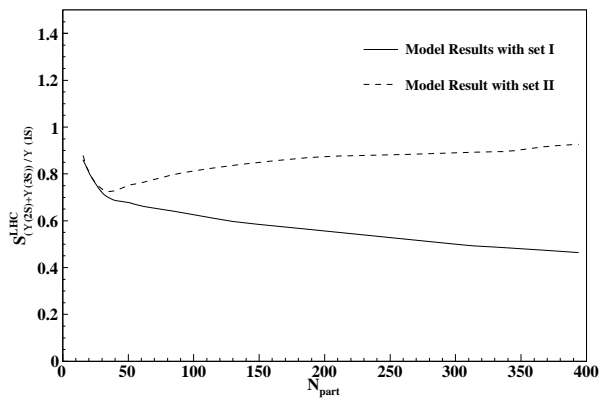}
    \label{fig:Fig 7}
    \caption{Prediction of variation of survival probability of $\Upsilon(2S)+\Upsilon(3S)/\Upsilon(1S)$ with respect to number of participants ($N_{part}$) using dissociation temperature of set I (solid curve) and set II (dashed-curve) at LHC energy ($\sqrt{s_{NN}}=2.76$ TeV).}
\end{figure}
\begin{figure}[!ht]
    \centering
        \includegraphics[height=18em]{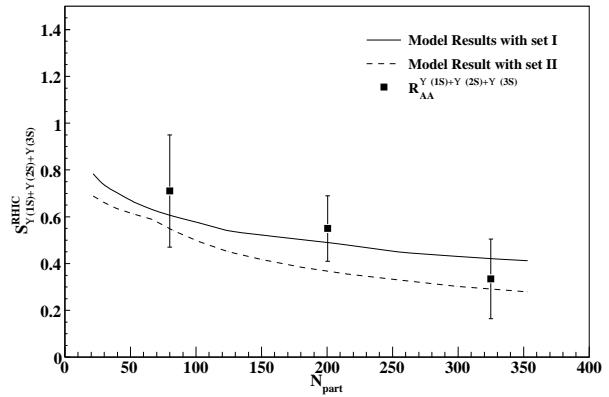}
    \label{fig:Fig 9}
    \caption{Variation of survival probability of $\Upsilon(1S)+\Upsilon(2S)+\Upsilon(3S)$ with respect to number of participants ($N_{part}$) using dissociation temperatures of set I (solid curve) and set II (dashed-curve) at highest RHIC energy ($\sqrt{s_{NN}}=200$ GeV). Solid squares are the experimental data points without CNM normalization~\cite{rosi}.}
\end{figure}
\begin{figure}[!ht]
    \centering
        \includegraphics[height=18em]{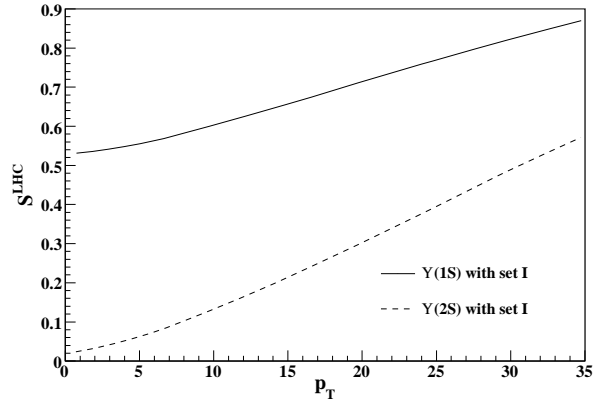}
    \label{fig:Fig 9}
    \caption{Variation of survival probability of $\Upsilon(1S)$ (solid curve) and $\Upsilon(2S)$ (dashed curve) with respect to $p_{T}$ for most central collision using dissociation temperature of set I at LHC energy ($\sqrt{s_{NN}}=2.76$ TeV).}
\end{figure}


\begin{figure}[!ht]
    \centering
        \includegraphics[height=18em]{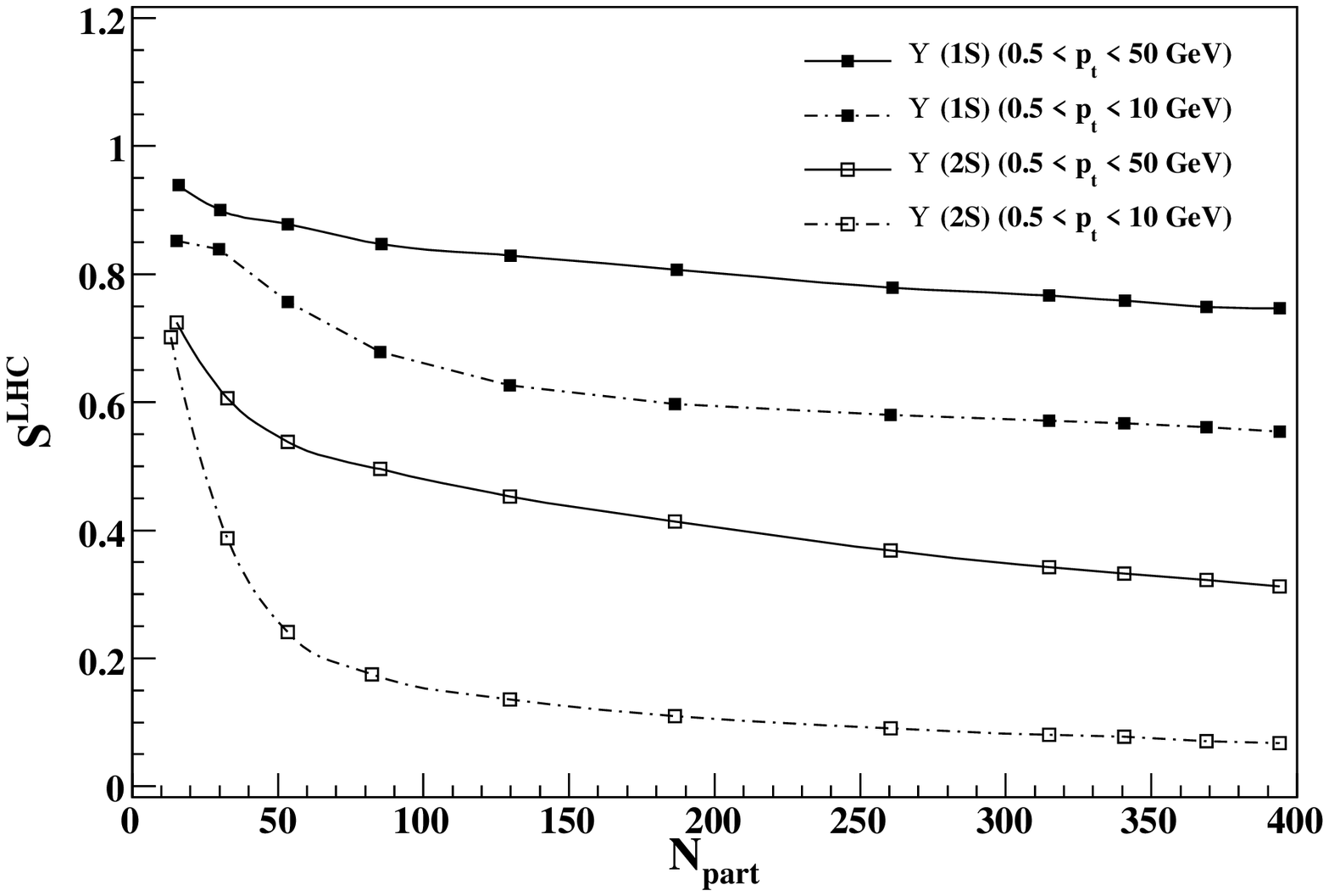}
    \label{fig:Fig 9}
    \caption{Variations of survival probabilities of $\Upsilon(1S)$ (solid curve) and $\Upsilon(2S)$ (dashed curve) with respect to $N_{part}$ using dissociation temperature of set I at LHC energy ($\sqrt{s_{NN}}=2.76$ TeV). Solid curve with solid square symbols and solid curve with hollow square symbols present the survival probabilities of $\Upsilon(1S)$ and $\Upsilon(2S)$, respectively in $0.5 ~< ~p_{T} ~<~50 $ GeV range. Dashed-dotted curve with solid square symbols and dashed-dotted curve with hollow square symbols show the $\Upsilon(1S)$ and $\Upsilon(2S)$, respectively in $0.5 ~< ~p_{T} ~<~10 $ GeV range. }
\end{figure}
\section{Results and Discussions}

Earlier investigations based on the potential models predict the dissociations of different bottomonia states at higher temperatures~\cite{satz}. However, recent calculation based on lattice QCD for free energy of heavy quarks finds the dissociation temperature somewhat lower than the earlier findings~\cite{mocsy}. Thus, there is uncertainity in the dissociation temperatures of upsilon and its various excited states. Exploiting this uncertainty, we find it worthwhile to investigate the effect of $T_{d}$ in simultaneously explaining the data of RHIC and LHC experiments. Thus we list two different sets of dissociation temperatures for bottomonia states ( labelled as set I and set II) as given in Table. 2.

 Fig. 1 shows the variations of $p_T$ integrated survival probability of $\Upsilon(1S)$ ($S_{\Upsilon(1S)}^{LHC}$) with centrality i.e., $N_{part}$ obtained from our present model using dissociation temperatures as given in set I. We also show comparison of our results with the nuclear modification factor ($R_{AA}$) of $\Upsilon(1S)$ obtained by CMS experiment~\cite{cms3}. Moreover, one should keep in mind that from here onwards, we use the same $p_{T}$ range in all our calculations as was used in the related experimental data. We find that our model results satisfy the $R_{AA}$ data of $\Upsilon(1S)$ for pheripheral collisions. However there is a difference between our results and experimental data for central collsions. This suggests that the effect of CNM is mild in pheripheral collisions and it increases with the centrality of the collision. In order to elucidate this point we determine the theoretical values of CNM factor ($R_{AA}^{CNM}=~R_{AA}^{exp}/S_{\Upsilon(1S)}^{LHC}$). We have plotted the values of $R_{AA}^{CNM}$ in fig. 1 and we believe this will provide a guideline for the experimentalists to test this prediction.

 Fig. 2 demonstrates the variation of $S_{\Upsilon(2S)}^{LHC}$ with respect to $N_{part}$ as obtained from our model with dissociation temperatures of set I. We also show comparison of our results with the $R_{AA}$ of $\Upsilon(2S)$ obtained by CMS experiment~\cite{cms3}. We again observe that the results are in fair agreement with the $R_{AA}$ data of $\Upsilon(2S)$ at least for pheripheral collisions. However, there is indication of CNM effect playing a role on the survival probability of $\Upsilon(2S)$ for central collsions. The overall suppression of $\Upsilon(2S)$ at all centralities is more than the $\Upsilon(1S)$ state as expected from the basic mechanism of Debye screening of colour charges. We also calculate and plot the nuclear modification factor ($R_{AA}^{CNM}$) as determined theoretically.

 Fig. 3 shows the variation of $S_{\Upsilon(2S)}^{LHC}/S_{\Upsilon(1S)}^{LHC}$ with respect to $N_{part}$ obtained in our model with set I. Our results are quite in agreement with the experimental data of $R_{AA}^{2S}/ R_{AA}^{1S}$ in pheripheral as well as most central collision. This confirms the finding that the ratio of the different upsilon state is very much free from CNM effects. However, a sizable difference in semi-central collisions outlines the importance of CNM on different upsilon states in these collisions. We have also plotted the variations of $S_{\Upsilon(2S)}^{LHC}/S_{\Upsilon(1S)}^{LHC}$ with respect to $N_{part}$ obtained in our model using dissociation temperatures of set II. The results obtained from set II show a large difference with the data and thus a large effect of CNM on $S_{\Upsilon(2S)}^{LHC}/S_{\Upsilon(1S)}^{LHC}$ is noticed. 



Fig. 4 shows the variation of the ratio $S_{\Upsilon(2S)}^{LHC}+S_{\Upsilon(3S)}^{LHC}/S_{\Upsilon(1S)}^{LHC}$ with respect to $N_{part}$ obtained in our model using dissociation temperatures as mentioned in set I and set II of table II. These results again suggest that the results with dissociation temperatures of set II show anomalous features in semi-central as well as central collisions. Thus set I is found to favourably describe the features of experimental data.

 In Fig. 5, we have presented our results for  $S_{\Upsilon(1S)+\Upsilon(2S)+\Upsilon(3S)}^{RHIC}$ with respect to $N_{part}$ using both set I and II. We also compare our results with the nuclear modification factor ($R_{AA}$) for $\Upsilon(1S)+\Upsilon(2S)+\Upsilon(3S)$ obtained by RHIC experiment~\cite{rosi}. Our results with the dissociation temperatures as given in both sets match well with the experimental data. However, experimental data still favour more the results with set I. This study suggests that the effect of CNM on the $\Upsilon(1S)+\Upsilon(2S)+\Upsilon(3S)$ at RHIC energy is very small and the suppression of $\Upsilon(1S)+\Upsilon(2S)+\Upsilon(3S)$ can be considered mainly due to sequential melting of upsilon states in QGP.

In Fig. 6, we show our results for the variation of the survival probabilities of $\Upsilon(1S)$ and $\Upsilon(2S)$ with respect to $p_{T}$ using dissociation temperatures given by set I. Fig. 6 then illustrates how different $p_{T}$ ranges can affect the survival probabilities of $\Upsilon$ states. We consider the most central collisions at the LHC energy in calculating the survival probabilities at various $p_{T}$. We have also incorporated the feeddown from higher states. The survival probabilities of $\Upsilon(1S)$ and $\Upsilon(2S)$ both start from a small value ie., $0.531$ and $0.023$, respectively at low $p_{T}$ and then increase with an increase in $p_{T}$ and get the values of $0.87$ and $0.573$ for $\Upsilon(1S)$ and $\Upsilon(2S)$, respectively at $p_{T}=35$ GeV. This again endorses the statement that upsilon states with large momentum can be formed at a later stage in the plasma rest frame. Consequently the region covered by a hot plasma is reduced and hence less suppression for such large momentum upsilon states~\cite{karsch}. In other words, more upsilon states at large momentum should survive.
 
As shown in Fig. 6, transverse momentum of $b-\bar{b}$ pair has significant impact on the survival probability of upsilon. Thus it is worthwhile to compare the survival probabilities of $\Upsilon (1S)$ and $\Upsilon (2S)$ calculated using different $p_{T}$-ranges. In Fig. 7, we present variations of survival probabilities of $\Upsilon (1S)$ and $\Upsilon (2S)$ with respect to $N_{part}$ in two different momentum range. First momentum range ie., $0.5~<~ p_{T}~<~50$ GeV is same as considered by CMS collaboration at LHC. We compare the survival probabilities of $\Upsilon (1S)$ and $\Upsilon (2S)$ in this momentum range with the survival probabilities calculated in a low $p_{T}$ range ie., for $0.5~<~ p_{T}~<~10$ GeV. These survival probabilities calculated in a low $p_{T}$ range might be measurable at detectors other than the CMS experiment in future. This study again confirms that upsilon states with large momentum are less suppressed in comparison to low $p_{T}$ upsilon states at the same centrality.

In summary, we have presented here a modified colour-screening model for precise determination of the suppression of various $\Upsilon$ states in QGP medium where we have used a quasi-particle model (QPM) equation of state for QGP and feed down from higher resonance states (namely, $\Upsilon(2S)$, $\Upsilon(3S)$, $\chi_{b1}$, and $\chi_{b2}$). We have further used the concept of dilated formation time for quarkonia states under consideration and viscous effects of the QGP medium was additionally used. We find that the studies of interactions of heavy quarks such as charm and bottom with QGP play a dominant role in enhancing our understanding regarding the properties of the QGP medium. Since these heavy quarks are dominantly produced by gluon-fusion in the early stages of the collision, they naturally experience the complete evolution of the system. We find that the mechanism of the modified colour screening proposed here explains the charmonia and bottomonia suppressions in the QGP medium in a unified way in the entire wide range of energy. Thus we conclude that the theoretical studies regarding quarkonia ($J/\psi$ and $\Upsilon$) productions and their quantitative comparisons with the existing experimental data obtained for different colliding systems, collision energies and centralities can be used as a guiding factor in disentangling the interplay of various mechanism regarding the effects of the medium properties.  
\noindent
\section{Acknowledgments}
PKS and SKT acknowledge the University Grant Commission (UGC) and Council of Scientific and Industrial Research (CSIR), New Delhi for financial support.

\newpage

\end{document}